\documentclass[twocolumn,showpacs,aps,amsmath,amssymb]{revtex4}
\usepackage{graphicx}
\usepackage{dcolumn}
\usepackage{bm}
\usepackage{amssymb}

\begin{document}

\title{Synchronization of Excitatory Neurons with Strongly Heterogeneous Phase Responses}

\author{Yasuhiro Tsubo}
\email{tsubo@brain.riken.jp}
\author{Jun-nosuke Teramae}
\author{Tomoki Fukai}
\affiliation{Laboratory for Neural Circuit Theory,
 RIKEN Brain Science Institute, Saitama 351-0198, Japan}

\date{\today}

\begin{abstract} 
   In many real-world oscillator systems, the phase response curves are
 highly heterogeneous.
   However, dynamics of heterogeneous oscillator networks has not been
 seriously addressed.
   We propose a theoretical framework to analyze such a system by
 dealing explicitly with the heterogeneous phase response curves.
   We develop a novel method to solve the self-consistent equations
 for order parameters by using formal complex-valued phase variables,
 and apply our theory to networks of in vitro cortical neurons.
   We find a novel state transition that is not observed in previous
 oscillator network models.
\end{abstract}


\maketitle

\begin{figure}[htbp]
\includegraphics[width=86mm]{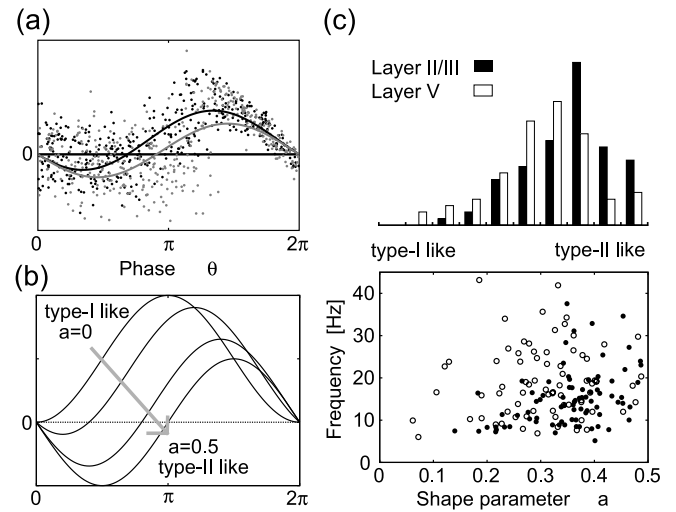}
\caption{
   Heterogeneity of the phase response curves (PRCs) of cortical neurons
 in our \textit{in vitro} recording studies \cite{Tsubo07}.
   We performed the least-square-error fitting of the PRCs with
 $Z(\theta)=-\cos(\theta-a\pi)+\cos a\pi$ by changing the shape
 parameter $a$ (see also Eq. (\ref{eqn-concrete_model})).
   (a) Two typical examples of the estimated PRCs.
   (b) The dependence of the PRC shape on $a$.
   (c) The distributions of the estimated shape parameters (upper)
 and the intrinsic frequencies of neuronal oscillators (lower)
 in different cortical layers.
   The frequency was tuned in experiments by varying the amplitude
 of a DC injected current.
}
\label{Figure1}
\end{figure}

   Synchronization phenomena are ubiquitous in nonlinear dynamical
 systems, such as Josephson junction arrays \cite{JosephsonJA},
 laser arrays \cite{Laser} and biological systems \cite{Winfree,
 Winfree67,BrainSci}.
   Synchronous firing of cortical neurons is considered to play
 an active role in cognitive functions \cite{BrainSci}, and is
 governed by the intrinsic properties of neurons as well as by
 the network connectivity.
   These properties include the phase response curve 
 (PRC) \cite{Winfree67, Winfree, Kuramoto84}, which describes
 how the timing of a succeeding output spike is shifted by an
 input spike \cite{Reyes93, Hansel95, Ermentrout96, Galan05, Preyer05}.
   In general, we can categorize the phase responses of
 cortical neurons into two types.
   Type-I PRC has only positive values (corresponding to phase
 advances), while type II has both positive and negative values
 (corresponding to phase delays) depending on the phase at which
 a stimulus is applied \cite{Hansel95}.
   Mutual synchronization of excitatory neurons may be easier
 with type-II PRC than with type-I PRC, if the PRCs of the
 neurons are homogeneous \cite{Hansel95}.
   However, the PRCs recorded from various brain areas, which
 include the hippocampus \cite{Lengyel05}, the entorhinal
 cortex \cite{Netoff05}, the somatosensory cortex \cite{Tateno07}
 and the motor cortex \cite{Reyes93, Tsubo07}, have revealed that
 the PRC type of pyramidal neurons is highly heterogeneous,
 especially if they belong to different cortical layers \cite{Tsubo07}.
   Even if two neurons have the same PRC type, the shape of PRC
 varies significantly from neuron to neuron (see Fig. \ref{Figure1}).

   In general, the heterogeneity of PRCs disturbs the stability
 of the synchronous state.
   However, the heterogeneity
 of PRCs can be compensated by other intrinsic
 properties that enhance synchronization.
   In this paper, we explore how such compensation may occur 
in networks of heterogeneous oscillators.
   The population dynamics of the heterogeneous phase oscillators
 was first studied in the Kuramoto model \cite{Kuramoto84}, which
 demonstrated the emergence of transitions between synchronized
 and desynchronized states \cite{Sakaguchi86, Daido01, Teramae04,
 Shinomoto06}.
   These studies transformed the heterogeneity of the PRC shapes
 into that of the natural frequencies, assuming that the heterogeneity
 is weak in both cases.
   Here, we develop an analytical method to explicitly deal with
 the heterogeneity since the PRC shapes of cortical neural oscillators
 are strongly heterogeneous.
   Unlike in the original Kuramoto model, we can show that the 
 order parameter of synchronization changes discontinuously 
 in the neural population with heterogeneous PRCs.
   This implies that this type of heterogeneity creates a dramatic
 effect on networks of neural oscillators.


   We first derive the interaction functions 
 $\Gamma_{j}(\psi)$ \cite{Kuramoto84} from a set of differential
 equations of the phase oscillators with a common frequency $\omega$.
   These oscillators have heterogeneous PRCs, $Z_{j}(\theta)$,
 and are globally coupled with each other.
   The phase of the $j$th oscillator $\theta_{j}$ obeys the evolution equation,
\begin{equation}
\frac{d\theta_{j}}{dt}=\omega
+\frac{\epsilon}{N}Z_{j}(\theta_{j})
\sum_{k=1}^{N}\sum_{n}\alpha(t-t_{k}^{n}).
\label{eqn-dynamics_theta}
\end{equation}
   The input to the $j$th oscillator from the $k$th is given as
 $\epsilon N^{-1}\sum_{n}\alpha(t-t_{k}^{n})$, where $\epsilon$ is a
 weak coupling constant, $N$ the number of oscillators, $\alpha(t)$
 a causal coupling function, and the $n$th firing time of the $k$th
 oscillator $t_{k}^{n}$ defined as $\theta_{k}(t_{k}^{n})=2\pi n$.

   The mutual interactions shift the frequency of the mean phase of
 these oscillators by $\epsilon\Omega$ from the natural frequency
 $\omega$.
   We define the relative phase $\psi_{j}=\theta_{j}-\Phi$, where
 the phase $\Phi=(\omega+\epsilon\Omega) t$.
   The relative phase $\psi_{j}$ changes slowly compared with
 $\theta_{j}$ and will hardly change during the oscillation period
 $2\pi/(\omega+\epsilon\Omega)$.
   Therefore, we can average Eq. (\ref{eqn-dynamics_theta}) over
 one period keeping $\psi_{j}$ constant.
   Using $(\omega+\epsilon\Omega) t_{k}^{n}=2\pi n-\psi_{k}$,
 we can describe  the dynamics of the relative phase $\psi_{j}$ as
\begin{eqnarray}
\frac{1}{\epsilon}\frac{d\psi_{j}}{dt}=-\Omega+
\frac{1}{2\pi N}\int_{0}^{2\pi}
\hspace*{-7pt}
Z_{j}(\psi_{j}+\Phi)\sum_{k}
\beta(\psi_{k}+\Phi) d\Phi,
\nonumber \\
\label{eqn-dynamics_psi}
\end{eqnarray}
   where $\beta(\psi_{k}+\Phi)$ is the sum of the contributions of
 past inputs from the $k$th oscillator
 $\sum_{n} \alpha((\psi_{k}+\Phi-2\pi n)/(\omega+\epsilon\Omega))$.
   In the limit of $N\rightarrow\infty$, we can apply the mean-field
 approximation, $N^{-1}\sum_{k}\beta(\psi_{k}+\Phi)\sim
 \int_{0}^{2\pi}\beta(\psi+\Phi)P(\psi)d\psi$, 
 where $P(\psi)$ is the distribution function of the relative phase
 $\psi$.
   The functions $Z,~P$ and $\beta$ are $2\pi$-periodic, so we can
 expand them into Fourier series.
   The $n$th Fourier coefficient $r_{n}^{(f)}e^{i\lambda_{n}^{(f)}}$
 of a function $f(\psi)$ is defined as $r_{n}^{(f)}e^{i\lambda_{n}^{(f)}}
 \equiv(2\pi)^{-1}\int_{0}^{2\pi}f(x)e^{-inx}dx$.
   Then, Eq. (\ref{eqn-dynamics_psi}) becomes,
\begin{eqnarray}
\frac{1}{\epsilon}\frac{d\psi_{j}}{dt}
&\hspace*{-1mm}=&
\hspace*{-1mm}
\Gamma_{j}(\psi_{j})
\label{eqn-dynamics_gamma} \\
&\hspace*{-1mm}\equiv&
\hspace*{-1mm}
\tilde{\omega}_{j}-\Omega+
\sum_{n=1}^{\infty}
K_{j}^{n}R^{n}\cos\left(n\psi_{j}+\Delta_{j}^{n}+\lambda_{n}^{(P)}\right),
\nonumber
\end{eqnarray}
   where $\tilde{\omega}_{j}=r_{0}^{(Z_{j})}r_{0}^{(\beta)}$,
 $K_{j}^{n}=2r_{n}^{(Z_{j})}r_{n}^{(\beta)}$
 and $\Delta_{j}^{n}=\lambda_{n}^{(Z_{j})}-\lambda_{n}^{(\beta)}$.
   The order parameters $R^{n}=2\pi r_{n}^{(P)},~\Omega$, 
 and $\lambda_{n}^{(P)}$ in Eq. (\ref{eqn-dynamics_gamma}) describe
 the dynamics of neural oscillators.

   Below, we focus on non-trivial solutions ($R^{n}\neq0$) to
 Eq. (\ref{eqn-dynamics_gamma}) other than the trivial one $R^{n}=0$.
   Each oscillator exhibits two dynamical modes, `synchronized' or
 `desynchronized', according to the shape of the PRC or the
 interaction function $\Gamma_{j}(\psi)$.
   In the `synchronized' mode, the oscillator is trapped at a
 stable fixed point of Eq. (\ref{eqn-dynamics_gamma}).
   In the `desynchronized' mode, Eq. (\ref{eqn-dynamics_gamma})
 has no stable fixed point, so the oscillator cannot be locked
 at any relative phase and drifts at a period of
 $T_{j}=\epsilon^{-1}\int_{0}^{2\pi}|\Gamma_{j}(\psi)|^{-1}d\psi$.
   To derive the order-parameter equations, we introduce complex
 order parameters $R^{n}e^{i\lambda_{n}^{(P)}}$, and divide them
 into the contributions of the `synchronized' population,
 ${\cal O}_{\rm s}^{n}$, and those of the `desynchronized' population,
 ${\cal O}_{\rm ds}^{n}$.
   Given the phase distribution function of the `synchronized'
 oscillators, $P_{\rm s}$, we can represent ${\cal O}_{\rm s}^{n}$ as 
\begin{equation}
{\cal O}_{\rm s}^{n}\equiv\int_{0}^{2\pi}
P_{\rm s}(\psi)e^{-in\psi} d\psi
=\frac{1}{N}\sum_{j~\in D}e^{-in\psi^{\dagger}_{j}},
\label{eqn-order_para_sync}
\end{equation}
   where $D$ refers to the indices of the `synchronized' oscillators
 and $\psi_{j}^{\dagger}$ is a real-valued solution to
 $\Gamma(\psi_{j}^{\dagger})=0$ satisfying
 $\Gamma'(\psi_{j}^{\dagger})<0$.
   In general, the equation has more than one solution.
   However, Eq. (\ref{eqn-order_para_sync}) can uniquely be defined
 in the limit of weak noise since the noise excludes solutions other
 than the most stable one.

   The contribution of the `desynchronized' oscillators
 ${\cal O}_{\rm ds}^{n}$ is given as
\begin{eqnarray}
{\cal O}_{\rm ds}^{n}&\equiv&\int_{0}^{2\pi}
P_{\rm ds}(\psi)e^{-in\psi} d\psi
\nonumber \\
&=&
\frac{1}{N}\sum_{j~\in \bar{D}}~
\int_{0}^{2\pi}d\psi
\frac{1}
{T_{j}|\Gamma_{j}(\psi)|}e^{-in\psi},
\label{eqn-order_para_desync}
\end{eqnarray}
   where $\bar{D}$ is the set of the indices of the `desynchronized'
 oscillators and $P_{\rm ds}$ is their phase distribution function.
   We can calculate the above integral using the residue theory as
\begin{eqnarray}
{\cal O}_{\rm ds}^{n}
&=&\frac{1}{N}\sum_{j~\in \bar{D}}~e^{-in\psi_{j}^{\dagger}}.
\label{eqn-order_para_desync_residue}
\end{eqnarray}
   Here, $e^{-in\psi_{j}^{\dagger}}=\left<e^{-in\psi^{(k)}_{j}};(\Gamma'(\psi_{j}^{(k)}))^{-1}\right>_{k}$,
 with $\psi_{j}^{(k)}$ being an imaginary solution to
 $\Gamma(\psi_{j}^{(k)})=0$ satisfying ${\rm Im}~\psi_{j}^{(k)}<0$
 and the weighted average $\left<f(x_{k});g(x_{k})\right>_k$ defined as 
 $\left<f(x_{k});g(x_{k})\right>_k\equiv\sum_{k}g(x_{k})f(x_{k})/\sum_{k}g(x_{k})$.
   We find that the formal expressions of the contributions of the
 `desynchronized' and `synchronized' populations are identical.
   Thus, we finally obtain the following self-consistent equation:
\begin{eqnarray}
R^{n}e^{i\lambda_{n}^{(P)}}&=&\frac{1}{N}\sum_{j}~e^{-in\psi^{\dagger}_{j}}.
\label{eqn-self_consistent}
\end{eqnarray}
   This equation means that the $n$th complex order parameters
 $R^{n}e^{i\lambda_{n}^{(P)}}$ should be identical with the
 $n$th circular moment of the complex solutions to $\Gamma_{j}(\psi)=0$.

   To obtain the explicit formula for the fixed points, we hereafter
 truncate $\Gamma(\psi)$ up to the first Fourier mode of
 Eq. (\ref{eqn-dynamics_gamma}).
   Then, the complex solutions $\psi^{\dagger}$ are given by
\begin{equation}
e^{-i\psi^{\dagger}_{j}}=
\left\{
\begin{array}{lc}
\left(-W_{j}+\sqrt{W^2_{j}-1}\right)e^{i\Delta^{1}_{j}}
& (W_{j}\geqq1)
\\
\left(-W_{j}-\sqrt{W^2_{j}-1}\right)e^{i\Delta^{1}_{j}}
& (W_{j}<1)
\end{array}
\right.,
\label{eqn-complex_solution}
\end{equation}
   where $W_{j}\equiv (\tilde{\omega}_{j}-\Omega)/K_{j}^{1}R^{1}$.
 If a single variable $a$ parameterizes the heterogeneity of the
 PRC shapes, we can explicitly describe the phase distribution functions as
\begin{eqnarray}
P_{\rm s}(\psi)&=&
g(a(\psi))\left|
\frac{\partial \log W}{\partial a}\frac{W}{\sqrt{1-W^2}}
-
\frac{\partial \Delta^{1}}{\partial a}
\right|^{-1},
\label{eqn-PDF_sync}
\\
P_{\rm ds}(\psi)&=&\frac{1}{2\pi}\int_{D_{\rm ds}}\frac{g(a)\sqrt{W^2-1}}{|W+\cos(\psi+\Delta^{1})|}da.
\label{eqn-PDF_nosync}
\end{eqnarray}
   where $g(a)$ is the distribution of the parameter values over
 the oscillator population.
   As mentioned previously, $\Gamma(\psi,a)=0$ has no real solution
 in the parameter range $D_{\rm ds}$.
   It is noted that merely finding the equilibrium values of the order
 parameters does not require the explicit expressions of
 $P_{\rm s, ds}(\psi)$ in the present analysis.

   To show the validity of our theoretical treatment and to get a novel
 insight into the dynamics of heterogeneous oscillator networks,
 we now apply it to coupled oscillators having the following 
 heterogeneous PRCs:
\begin{eqnarray}
Z(\theta)=-\cos(\theta-a\pi)+\cos a\pi.
\label{eqn-concrete_model}
\end{eqnarray}
   The value of the shape parameter $a$ is distributed uniformly
 in the range $(a_{\rm min}\leqq a \leqq a_{\rm max})$.
   A large or a small value of $a$ corresponds to type II- or 
 type I-like PRC, respectively.
   The coupling function $\alpha(\psi)$ is an exponential function
 with a decay constant of $\tau$: $\alpha(t)=\tau^{-1}\Theta(t)e^{-t/\tau}$,
 where $\Theta(x)$ is Heaviside function:
 $\Theta(x)=1$ if $x>0$ or 0 if $x<0$.
   Neurons synchronize easier with type II-like PRCs than with
 type I-like PRCs.
   Using Eq. (\ref{eqn-concrete_model}), we can rewrite
 Eq. (\ref{eqn-dynamics_gamma}) as
\begin{eqnarray}
\frac{1}{\epsilon K_{j}^{1}R^{1}}\frac{d\psi_{j}}{dt}
&=&W(a)-\cos\left(\psi_{j}-a\pi+{\rm Arctan}(\tau\omega)\right)
\nonumber \\
W(a)&=&\frac{\cos a\pi-2\pi\Omega/\omega}{R^{1}/\sqrt{1+(\tau\omega)^2}}
\label{eqn-concrete_dynamics}
\end{eqnarray}
   where $K_{i}^{1}=\omega/(2\pi\sqrt{1+(\tau\omega)^2})$.

\begin{figure}[htbp]
\includegraphics[width=86mm]{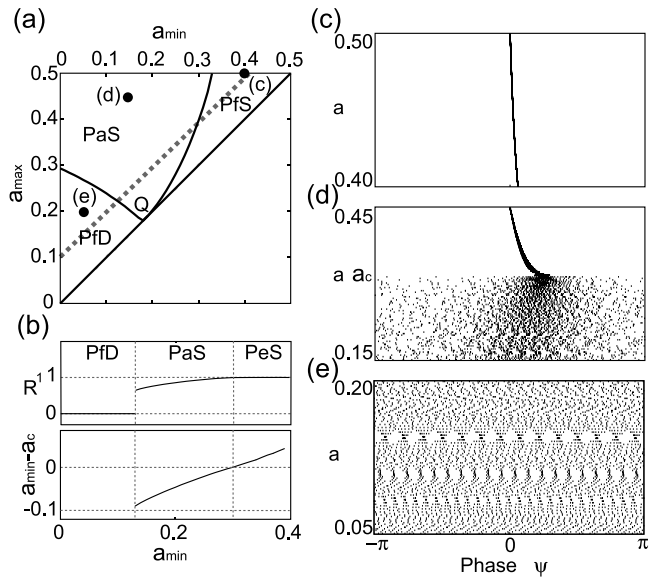}
\caption{
   Three different states in our coupled oscillator model with the
 PRCs represented by Eq. (\ref{eqn-concrete_model}).
   The parameters were set as $\tau=0.005,~\omega=40\pi,$ and
 $\epsilon=0.01$.
   (a) Phase diagram of this model, where the shape parameter $a$
 is distributed uniformly in $a_{\rm min}\leqq a \leqq a_{\rm max}$.
   The abbreviations mean perfectly synchronized (PfS), partially
 synchronized (PaS), and perfectly desynchronized (PfD) states.
   (b) The order parameters $R^{1}$ and $a_{\rm min}-a_{\rm c}$ are
 shown along the line $a_{\rm max}-a_{\rm min}=0.1$ (dashed line in
 (a)).
   The PfD state corresponds to $R^{1}=0$, while the PaS or the PfD
 state is defined with $R^{1}\neq 0$ and with $a_{\rm min}-a_{\rm c}<0$
 or $a_{\rm min}-a_{\rm c}>0$, respectively.
   (c)$-$(e) Raster plots of the relative phases $\psi_{j}$ of
 $100$ neural oscillators in the PfS (c), PaS (d), and PfD (e) states.}
\label{Figure2}
\end{figure}

   Using these results, we can study different dynamical states
 of the oscillator network.
   Figure \ref{Figure2}(a) summarizes the phase diagram in the
 ($a_{\rm min}$, $a_{\rm max}$) half plane.
   When Eq. (\ref{eqn-self_consistent_1st_hetero}) has a non-trivial
 solution, neurons are either partially synchronized (PaS) or
 perfectly synchronized (PfS).
   The border between the two states by can be determined by a
 critical value $a_{\rm c}$, which is a solution to $|W(a_{\rm c})|=1$.
   Since neurons with $a>a_{\rm c}$ are synchronized and those
 with $a<a_{\rm c}$ are desynchronized, the PfS state requires
 $a_{\rm min}>a_{\rm c}$.
   Otherwise, neurons are only partially synchronized.
   If $a_{\rm min}$ is sufficiently large, all neurons may be
 type II-like.
   We note that even in such a case, a strong heterogeneity
 (i.e., a sufficiently large $a_{\rm max}-a_{\rm min}$) may disable
 the perfect synchronization of oscillators. 

   The entire population of oscillators is perfectly desynchronized
 (PfD) if the self-consistent equation,
\begin{equation}
R^{1}=\frac{1}{N}\sum_{j}e^{-j\psi_{j}^{\dagger}}
\label{eqn-self_consistent_1st_hetero}
\end{equation}
 has no non-trivial ($R^{1}\neq 0$) solutions in the allowed range
 of $a$-value.
   Thus, the border between the PaS and PfD states is determined
 from the condition that a solution with $R^{1}\neq0$ exists.
   At $a_{\rm min}=a_{\rm max}$, the self-consistent equation has
 a stable fixed point if $a>\pi^{-1}{\rm Arctan}(\tau\omega)$ and
 the system exhibits the PfS state (point Q).
   The PfS-PaS and PaS-PfD borders should merge at this point.
   Figure \ref{Figure2}(c)-(e) displays the raster plots of the
 neural oscillators in the PfS, PaS, and PfD states designated
 in Figure \ref{Figure2}(a), respectively.
   In Figure \ref{Figure2}(c) or (e), the population comprises
 only type II-like or type I-like neurons showing perfect
 synchronization or perfect desynchronization, respectively.
   In Figure \ref{Figure2}(d), only sub-population of strongly
 type II-like neurons with $a>a_{c}$ are synchronized.

   The necessary condition for getting a non-trivial solution to
 Eq. (\ref{eqn-self_consistent_1st_hetero}) is that the imaginary part
 of its right hand should vanish.
   While the original Kuramoto model always satisfies this condition
 due to its symmetry, this is not the case in the present example.
   Therefore, this model and the Kuramoto model exhibit qualitatively
 different transitions from the PfD to the PaS state.
   In this model, the order parameter $R^{1}$, which vanishes in
 the PfD state, jumps discontinuously to a non-zero value at the
 transition point (Fig. \ref{Figure2}(b)).
   In contrast, the transition is continuous in the Kuramoto model.
   Therefore, the network state switches very sharply in the present
 model.
   Our quantitative results reveal that the different types of the
 heterogeneity can result in qualitatively different phase
 transition-like behaviors of oscillator networks.

\begin{figure}
\includegraphics[width=86mm]{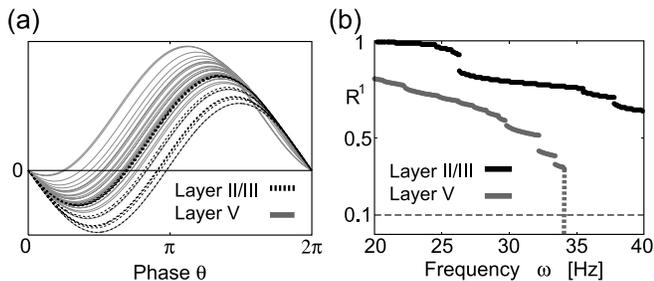}
\caption{
   The layer dependences of the PRCs of cortical neurons in the
 $\gamma$-frequency range and the order parameter $R^{1}$.
   (a) The PRCs recorded in \cite{Tsubo07} were fitted by Eq.
 (\ref{eqn-concrete_model}).
   PRCs were superimposed for 10 layer-II/III neurons
 (broken black lines) and 25 layer-V neurons (solid gray lines).
   (b) The order parameter $R^{1}$ depends differentially on the
 frequency $\omega$ of neuronal firing for layer-II/III (black)
 and layer-V (gray) neurons.
   Here, $\tau=0.005$.
   Eq. (\ref{eqn-self_consistent_1st_hetero}) in general has
 multiple solutions when the summation is taken over a finite number
 of oscillators.
   Here, we plotted the largest $R^{1}$, or plotted nothing
 if $R^{1}<0.1$.
   Dashed line indicates a discontinuous transition of the
 network state.
   Other discontinuous points of the curves appeared due to
 the finite-size effect.}
\label{Figure3}
\end{figure}
   We apply our theory to the data recorded from excitatory neurons
 in cortical layers II/III and V \cite{Tsubo07}.
   The PRC shapes exhibited a remarkable layer dependence when the
 firing frequency $\omega$ is in a range of 20-45 Hz, which is
 within the $\gamma$-frequency range (Fig. \ref{Figure3}(a)).
   To demonstrate the effect of this layer dependence on the population
 dynamics of cortical neurons, we analyze coupled systems of
 layer-II/III or layer-V neurons separately.
   The synchronizing property in general depends on the frequency
 $\omega$ as well as the decay time constant of excitatory synapses.
   As $\omega$ is increased, layer-V neurons display an abrupt
 transition from the PaS to PfD state, as represented by a
 discontinuous jump in $R^{1}$ (dashed line).
   This transition can also be found in the simpler model
 shown previously in Fig. \ref{Figure2} with a uniform distrubution of $a$. 

   These results may have considerable implications for exploring
 computational functions of local cortical circuits.
   Furthermore, in most real-world oscillator systems,
 the phase response curves are highly heterogeneous,
 so our theoretical method can be naturally applied to these systems.
   Our results provide a way to make quantitative predictions
 about the dynamics of general heterogeneous-oscillator networks, 
 indicating the existence of a clear-cut border between
 the PaS and PfD states with a discontinuous jump 
 in the order parameter.


   We thank Y. Kuramoto, T. Mizuguchi, H. Nakao, G. Bi, H. Cateau
 and N. Masuda for fruitful discussions and valuable comments.
   This work has been supported in part by Grant-in-Aid for
 Scientific Research in Priority Areas (17022036) and for Young
 Scientists (B) (50384722) from the Japanese Ministry of Education,
 Culture, Sports, Science and Technology.


\end{document}